\documentclass[ prl, twocolumn, nofootinbib, floatfix,superscriptaddress]{revtex4-2}

\usepackage{notes2bib}
\usepackage{natbib}
\usepackage{gensymb}
\usepackage{amsthm}
\usepackage{amsmath}
\usepackage{color}
\usepackage{bm}
\usepackage{amsmath,amssymb}
\usepackage{amsfonts}
\usepackage{dsfont}
\usepackage{graphicx}
\usepackage{float}
\usepackage{subfig}
\usepackage{verbatim}
\usepackage{amsfonts}
\usepackage{bbold}
\usepackage{dcolumn}
\usepackage{bm}
\usepackage[dvipsnames]{xcolor}
\usepackage{listings}
\definecolor{blueprl}{RGB}{46,48,146}
\usepackage[pdftex,colorlinks=true,urlcolor=blueprl,citecolor=blueprl,linkcolor=blueprl]{hyperref}
\usepackage{times}
\usepackage{color}
\usepackage[dvipsnames]{xcolor}

\usepackage{physics}
\usepackage{caption}
\usepackage{subcaption}

\usepackage[mathscr]{euscript}
\usepackage{amsmath}
\usepackage{graphicx}
\usepackage{dcolumn}
\usepackage{bm}
\usepackage{epsfig}
\usepackage{amssymb,latexsym,mathrsfs}
\usepackage{graphicx}
\usepackage{color}
\usepackage{hyperref}
\usepackage{float}
\usepackage{diagbox}
\usepackage[normalem]{ulem}

\newcommand{\vt}{\vphantom{\frac{1}{2}}}
\newcommand{\mtx}{\mathbf{R}}
\definecolor{vividviolet}{rgb}{0.62, 0.0, 1.0}
\definecolor{amaranth}{rgb}{0.9, 0.17, 0.31}
\definecolor{palatinateblue}{rgb}{0.15, 0.23, 0.89}
\definecolor{brightpink}{rgb}{1.0, 0.0, 0.5}
\definecolor{cornflowerblue}{rgb}{0.39, 0.58, 0.93}
\definecolor{deepcarminepink}{rgb}{0.94, 0.19, 0.22}
\definecolor{radicalred}{rgb}{1.0, 0.21, 0.37}
\definecolor{blueblue}{RGB}{21,47,181}
\definecolor{greengreen}{RGB}{65,166,16}

\newcommand{\be}{\begin{equation}}
\newcommand{\ee}{\end{equation}}
\newcommand{\bs}{\begin{split}} 
\newcommand{\bea}{\begin{eqnarray}}
\newcommand{\eea}{\end{eqnarray}}
\newcommand{\non}{\nonumber }

\usepackage{mathtools}
\newcommand{\D}{\mathrm{d}}
\newsavebox{\myhbar}

\newcommand{\p}{\partial}

\newcommand{\mrm}[1]{\mathrm{#1}}

\newcommand{\h}[1]{\hat{#1}}

\makeatletter
\def\maketitle{
\@author@finish
\title@column\titleblock@produce
\suppressfloats[t]}
\makeatother

\begin{document}

%\title{(1+1)-dimensional Hawking radiation model for a Kerr-Newman black hole} 

\title{General relativistic particle trajectories via quantum mechanical weak values and the Schwarzschild-Alcubierre spacetime}
\author{Joshua Foo}
\affiliation{Department of Physics, Stevens Institute of Technology, Castle Point Terrace, Hoboken, New Jersey 07030, U.S.A.}
\author{Cameron Bellamy}
\affiliation{Centre for Quantum Computation \& Communication Technology, School of Mathematics \& Physics, The University of Queensland, St.~Lucia, Queensland, 4072, Australia}
\author{Timothy C. Ralph}
\affiliation{Centre for Quantum Computation \& Communication Technology, School of Mathematics \& Physics, The University of Queensland, St.~Lucia, Queensland, 4072, Australia}

\begin{abstract}
We show that the average trajectories of relativistic quantum particles in Schwarzschild spacetime, obtained via quantum mechanical weak measurements of momentum and energy, are equivalent to the predicted flow lines of probability current in curved spacetime quantum theory. We subsequently demonstrate that these trajectories correspond exactly to classical null geodesics in a hybrid Schwarzschild-Alcubierre spacetime. This threefold equivalence demonstrates how quantum theory in curved spacetime can be formulated via operationally-defined measurements, and that such a theory may be interpreted deterministically, in the spirit of hidden-variable models such as Bohmian mechanics, through the novel connection to an underlying ``guiding metric.''
\end{abstract} 

\date{\today} 

\maketitle 

The counterintuitive nature of quantum theory has led to numerous debates over the physical interpretation of its predictions and underlying mathematical structure \cite{omnesRevModPhys.64.339,cramerRevModPhys.58.647}. The distinction between different interpretations usually occurs at the level of \textit{epistemology} and \textit{ontology}--respectively, our knowledge and the reality of physical systems and their properties. For example in the textbook Copenhagen interpretation \cite{heisenberg1985descriptive,bohr1928quantum}, the wavefunction $\psi(t,\mathbf{x})$ provides statistical information about finding a particle at the spacetime point $(t,\mathbf{x})$ through the Born rule, $\rho(t,\mathbf{x} ) = | \psi(t,\mathbf{x} ) |^2$. The irreversible act of measurement, and the collapse of the wavefunction that accompanies it, only reveals information about the system through the classical outcome of this measurement. 

In contrast with the probabilistic Copenhagen interpretation, Bohmian mechanics posits that particles follow real, deterministic trajectories \cite{bohmPhysRev.85.166,bohm2006undivided,Bell_Aspect_2004,holland1995quantum,durr2009bohmian}. Indeed, the seminal work of Einstein, Podolsky, and Rosen that questioned the completeness of quantum theory was pivotal in the development of hidden-variable theories such as Bohmian mechanics \cite{einsteinPhysRev.47.777,bellPhysicsPhysiqueFizika.1.195}. Since an actual configuration of particles with definite positions and velocities is proposed to exist independent of observation, it is a hidden-variable theory that is also nonlocal, in accordance with Bell's theorem \cite{bellPhysicsPhysiqueFizika.1.195}. For an $N$-particle system, the wavefunction determines the evolution of each particle, the dynamics of which are mutually nonlocally dependent on the positions and velocities of all other particles at a given instant of time. By its construction as an interpretation, Bohmian mechanics reproduces the predictions of standard nonrelativistic quantum theory. 

Discussions concerning the place of deterministic interpretations of quantum theory were revitalised following the paper by Wiseman \cite{wiseman2007grounding}, who demonstrated that operational meaning can be assigned to the average trajectories of nonrelativistic particles through quantum mechanical weak measurements \cite{aharonov1988result}. Wiseman showed that these trajectories, coupled with the additional assumption of determinism, are precisely those predicted by the Bohmian framework. Weak measurements are implemented by only weakly coupling the system of interest to the measurement device but allow one to obtain the expected value of a quantum-mechanical observable to arbitrarily high precision when performed repeatedly on an identically prepared ensemble \cite{dresselRevModPhys.86.307}. Weak values generalise weak measurements by including a subsequent strong (projective) measurement and post-selection based on the outcome of this measurement. Wiseman showed that a quantum-mechanical velocity field $v(t,\mathbf{x}) = j_\mathrm{S}(t,\mathbf{x})/\rho_\mathrm{S}(t,\mathbf{x})$, where $j_\mathrm{S}(t,\mathbf{x}), \rho_\mathrm{S}(t,\mathbf{x})$ are respectively the Schr\"odinger conserved current and density, can be obtained through the time-derivative of a position weak value, which was subsequently demonstrated in key experiments by Kocsis et.\ al. \cite{kocsis2011observing} and Mahler et.\ al \cite{mahler2016experimental}. This framework was also recently extended to special relativistic dynamics in Refs.\ \cite{foo2022relativistic,fooPhysRevA.109.022229}.

In spite of these notable results, the question of whether a consistent formulation of deterministic quantum theory in curved spacetime can be made, is still debated \cite{durrdoi:10.1098/rspa.2013.0699}. Indeed, a common criticism of formulations such as Bohmian mechanics is its apparent incompatibility with general relativity, stemming in part from the underlying nonlocality of the theory \cite{durr1990realistic,Horton_2004,SheldonGoldstein_2003,bravermanPhysRevLett.110.060406}, and related issues such as the non-positive-definiteness of the Klein-Gordon current density for bosons \cite{Ghose_2001,BOHM1987321,berndlPhysRevA.53.2062,horton2002broglie,2000horton,Alkhateeb2022,Berry_2012}, which gives rise to tangent vectors and hence particle trajectories that can be directed toward the past. Many of these interpretive issues arise because they take as a starting point various theories of or modifications to relativistic quantum mechanics, rather than an operational framework that grounds phenomena in the measurement of physical observables.

Here we provide a consistent description of relativistic quantum-mechanical particles propagating in curved spacetime whose average trajectories can be obtained via locally-measured weak values of momentum and energy. This generalises Wiseman's nonrelativistic framework to a fully general relativistic setting, in particular grounding the trajectories in operationally-defined measurements that can in-principle be carried out in an experiment. We focus on radially propagating massless scalar particles in the Schwarzschild spacetime exhibiting single- and multi-particle interference. We show that the velocity field of the particles is manifestly Lorentz covariant, being constructed from the time and space components of the scalar Klein-Gordon conserved current vector. We refer to the particles as photons since the scalar theory is a good approximation to a full quantum electrodynamics (QED) framework in the regimes we consider. To draw a connection with a deterministic particle ontology, we demonstrate that the average weak value trajectories correspond exactly with null geodesics in a hybrid spacetime that interpolates between the Schwarzschild and Alcubierre metrics \cite{alcubierre1994warp}. This classical spacetime, which is a physically admissible solution to Einstein's equations, is nevertheless a function of the quantum-mechanical wavefunction, a property that raises interesting questions concerning the interplay of quantum mechanics and general relativity. These results taken together (i) demonstrate how quantum theory in curved spacetime can be formulated in terms of operationally-defined measurements, and (ii) through the postulate of an underlying ``guiding metric,'' be interpreted through a deterministic viewpoint, contrary to prior expectations. 

Let us start by deriving an expression for the weak value velocity of photons in the Schwarzschild spacetime, with line element $\D s^2 = - f(r) \D t^2 + f(r)^{-1} \D r^2 + r^2 \D \Omega^2$, where $f(r) = 1-2m/r$ and $\D t^2, \D r^2$, and $\D \Omega^2$ are the temporal, radial, and spherical components of the metric. The radial velocity in Schwarzschild coordinates is given by $v(t,r) = \D r/\D t$. It will be convenient for us to work with tortoise coordinates \cite{misner1973gravitation}, related to the Schwarszchild coordinate via $r^\star = r + 2m \ln | r/2m - 1 |$, such that 
\begin{align}
    v(t,r) &= f(r) \frac{\D r^\star}{\D t} = f(r) \frac{\D r^\star}{\D \tau}\frac{\D \tau}{\D t} \equiv f(r) \frac{p(r^\star)}{E(r^\star)} 
    \label{eq1}
\end{align}
% Equation (\ref{eq1}) can be further expressed in terms of components of the 4-momentum via, 
% \begin{align}
%     v(t,r) &= f(r) \frac{\D r^\star}{\D \tau} \frac{\D \tau}{\D t} \equiv f(r) \frac{p(r^\star)}{E(r^\star)}
%     \label{eq2}
% \end{align}
where $p(r^\star), E(r^\star)$ are the radial and time components of the 4-momentum respectively. Using Eq.\ (\ref{eq1}), we propose an operational method of constructing the average (radial) velocity of quantum-mechanical photons in the Schwarzschild metric:
\begin{align}
    v(t,r) &= f(r) \frac{\langle \hat p_w \rangle}{\langle \hat H_w \rangle}
    \label{eq3}
\end{align}
where the notation $\langle \hat A_w \rangle$ is used to denote the \textit{weak value} of the observable $\h A$, defined as \cite{dresselRevModPhys.86.307}
\begin{align}
    \langle \hat A_w \rangle \equiv \mrm{Re} \frac{\langle \phi | \h A | \psi(t) \rangle}{\langle \phi | \psi(t) \rangle} ,
    \label{eq4}
\end{align}
and $\h p, \h H$ are the momentum and Hamiltonian operators respectively. Here, $| \psi(t) \rangle$ is the initial state of the system and $| \phi \rangle$ is the final state onto which it is projected, while the denominator of Eq.\ (\ref{eq4}) accounts for the fact that $\langle \h A_w \rangle$ is an ensemble average over postselected outcomes. From Eq.\ (\ref{eq4}), it follows that 
\begin{align}
    v(t,r) &= f(r) \frac{\mrm{Re} \langle \phi | \hat p | \psi(t) \rangle}{\mrm{Re} \langle \phi | \hat H | \psi(t) \rangle} . 
    \label{eq5}
\end{align} 
We remark that Eq.\ (\ref{eq3}) leading to Eq.\ (\ref{eq5}) is general, in that it provides a prescription for observers to measure the average trajectories (velocity) of relativistic particles with arbitrary spin in the Schwarzschild spacetime through weak values of the associated $\h p, \h H$. To connect the velocity field to average trajectories in spacetime, we shall consider initial Gaussian wavepackets $f(k)$ in momentum space, yielding a superposition of the form $| \psi(t) \rangle = \int\D k \: f(k) e^{-i\h Ht} | k \rangle$ and postselection onto position eigenstates $| \phi \rangle = \int\D k \: e^{-ikr^\star} | k \rangle$ with respect to the tortoise coordinate $r^\star$. Operationally, this corresponds to a broadband detector performing position measurements in the reference frame of a local observer. While the $(t,r^\star)$ coordinates may not seem to be the natural choice for such an observer, they correspond to such local measurements with respect to proper time and distance, once one accounts for redshift in the frequency. The utility of $(t,r^\star)$ coordinates  will become apparent when connecting Eq.\ (\ref{eq5}) to the relativistic Klein-Gordon theory in the Schwarzschild spacetime.

To evaluate Eq.\ (\ref{eq5}) for photons in the Schwarzschild spacetime, we consider dynamics described by the scalar Klein-Gordon equation, $g^{\mu\nu} D_\mu D_\nu \psi = 0$, where $g^{\mu\nu}$ is the metric tensor and $D_\mu$ is the covariant derivative \cite{Birrell_Davies_1982}. The scalar theory accurately describes the relativistic properties of photons with the understanding that it is a simplification of full QED. In particular, within the high-frequency (short-wavelength) approximation we employ (the quantum optical limit, as we explain below), polarisation-dependent effects are negligible \cite{skrotskii1957influence,balazsPhysRev.110.236,plebanskiPhysRev.118.1396,masshoonPhysRevD.7.2807,deriglazovPhysRevD.104.025006}. In this limit, the spacetime curvature does not vary significantly across the spatial extent of the wavefunction, and therefore the radial solutions (of which we are interested in) reduce to plane waves. This means that the dynamics of the 4-component spin-1 theory are well-described by the time-component i.e.\ the scalar theory \cite{nikolic2019relativistic} Moreover, in the spherically symmetric Schwarzschild spacetime we consider, rotational effects leading to polarisation-angular momentum coupling, are absent \cite{oanceaPhysRevD.102.024075}.

\begin{figure*}[t]
 \subfloat{\includegraphics[width=0.3\linewidth]{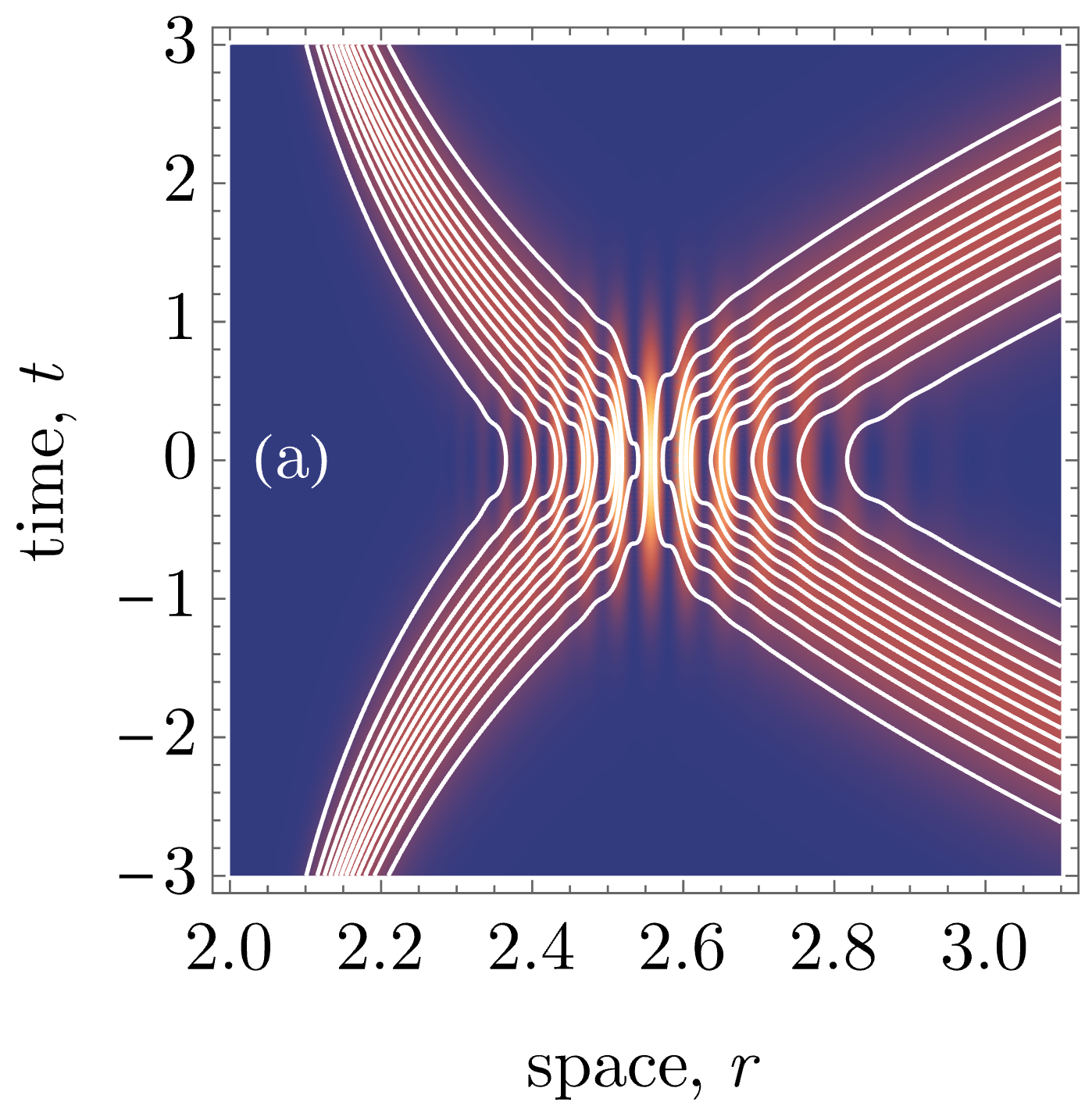}}\hspace{5pt}
 \subfloat{\includegraphics[width=0.3\linewidth]{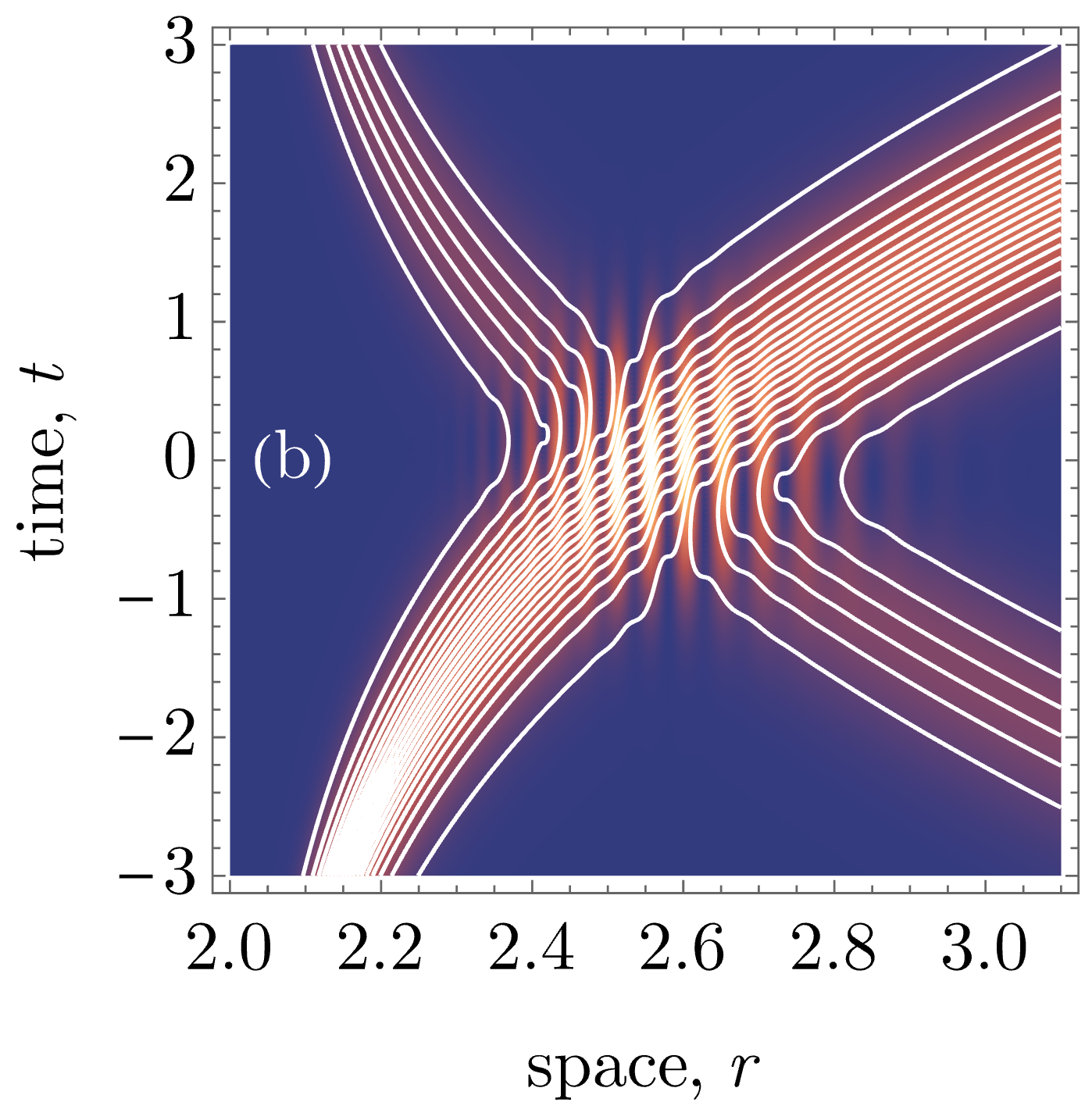}}\hspace{5pt}
 \subfloat{\includegraphics[width=0.3\linewidth]{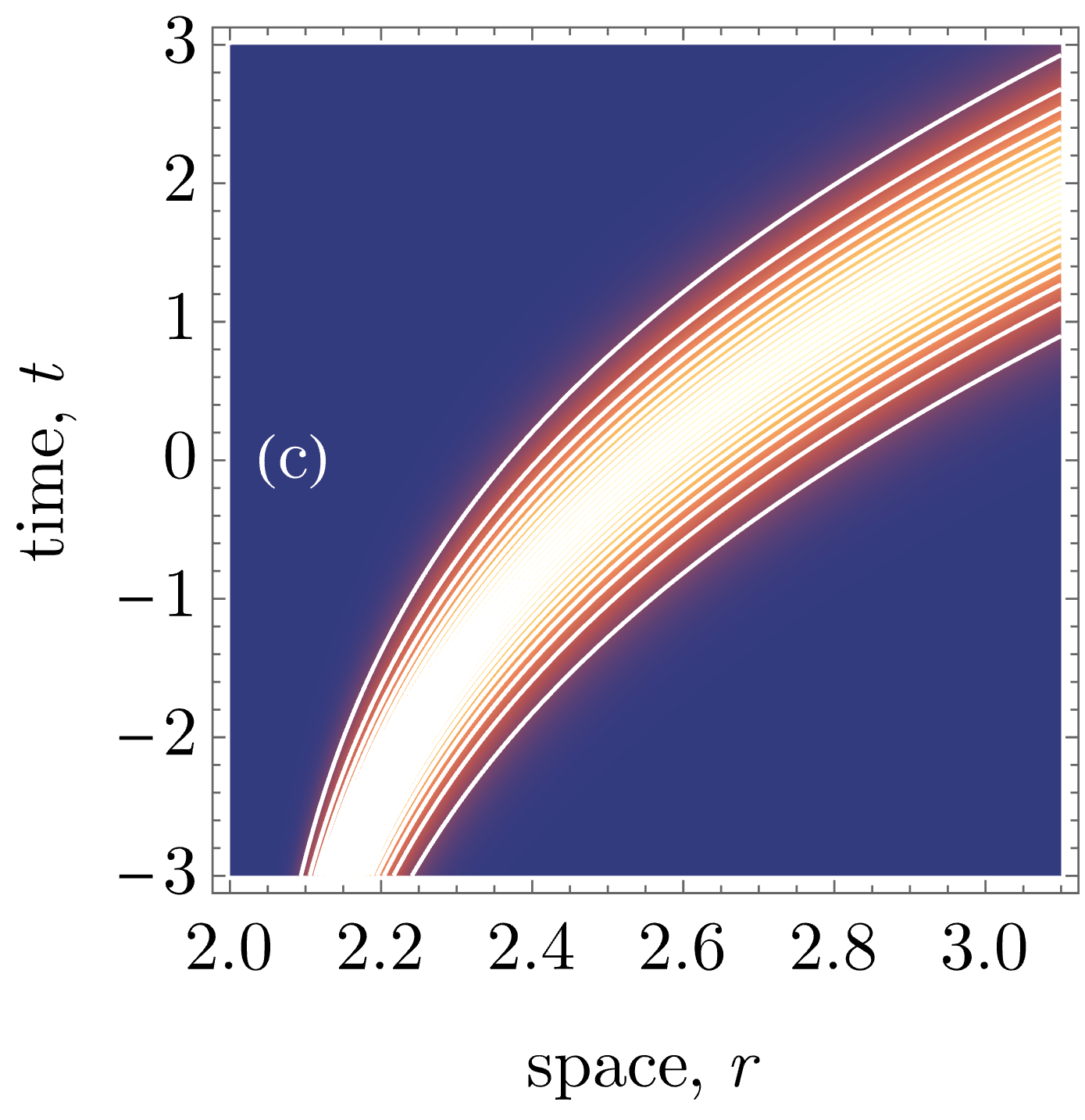}}\\[-2.5pt]  %%<-- in this line
 \subfloat{\includegraphics[width=0.3\linewidth]{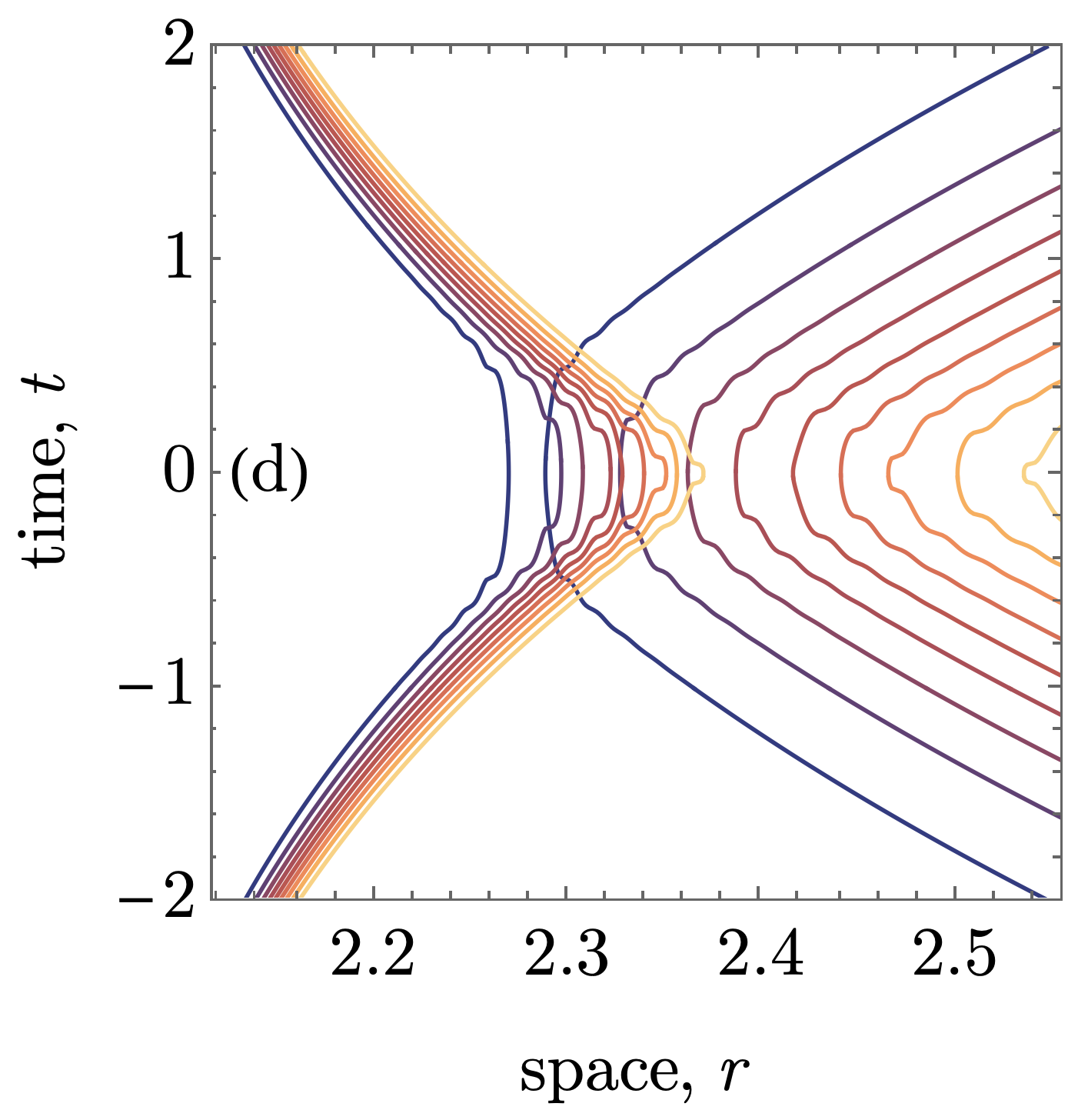}}\hspace{5pt}
 \subfloat{\includegraphics[width=0.3\linewidth]{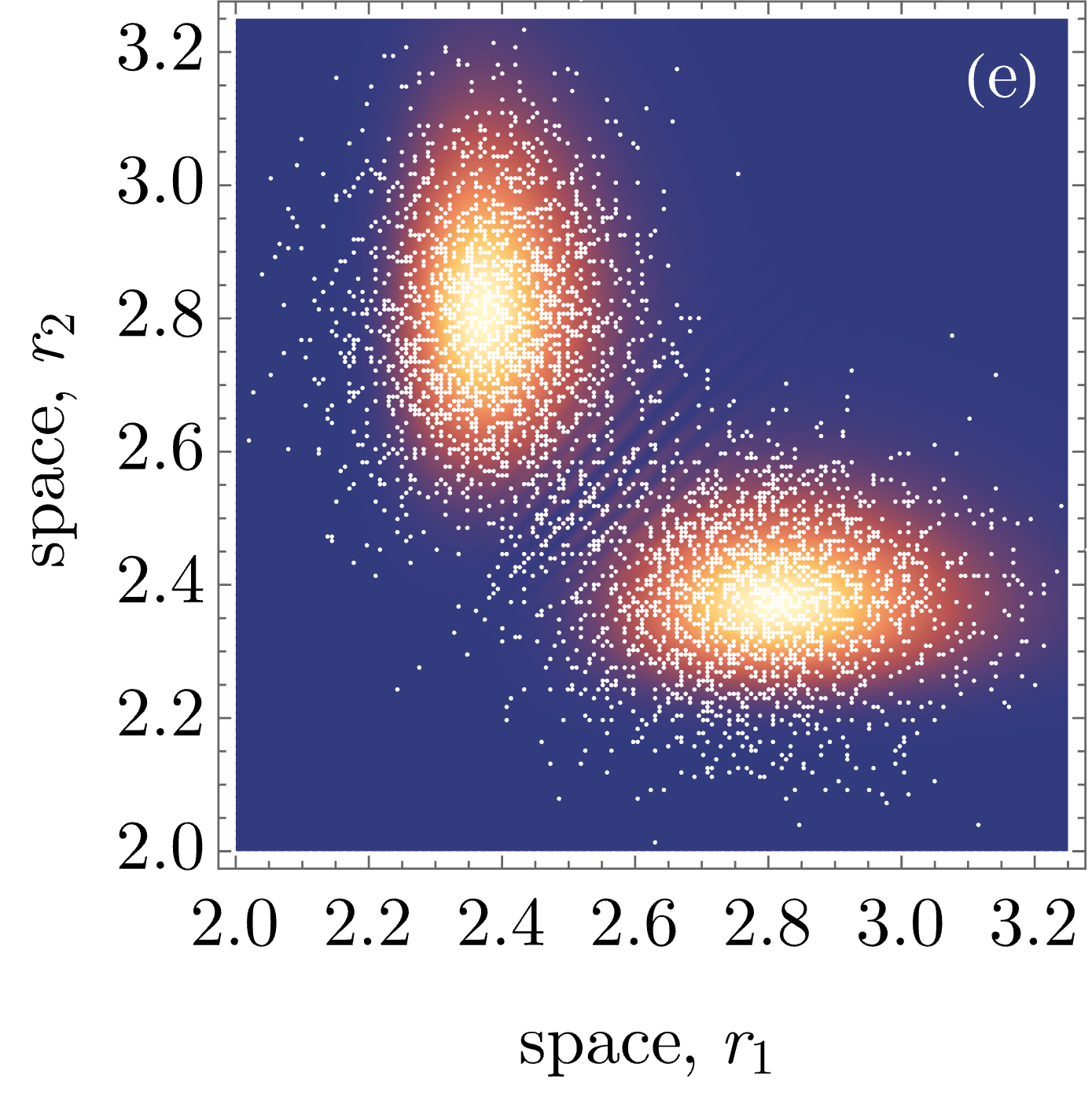}}
 \hspace{5pt}
 \subfloat{\includegraphics[width=0.3\linewidth]{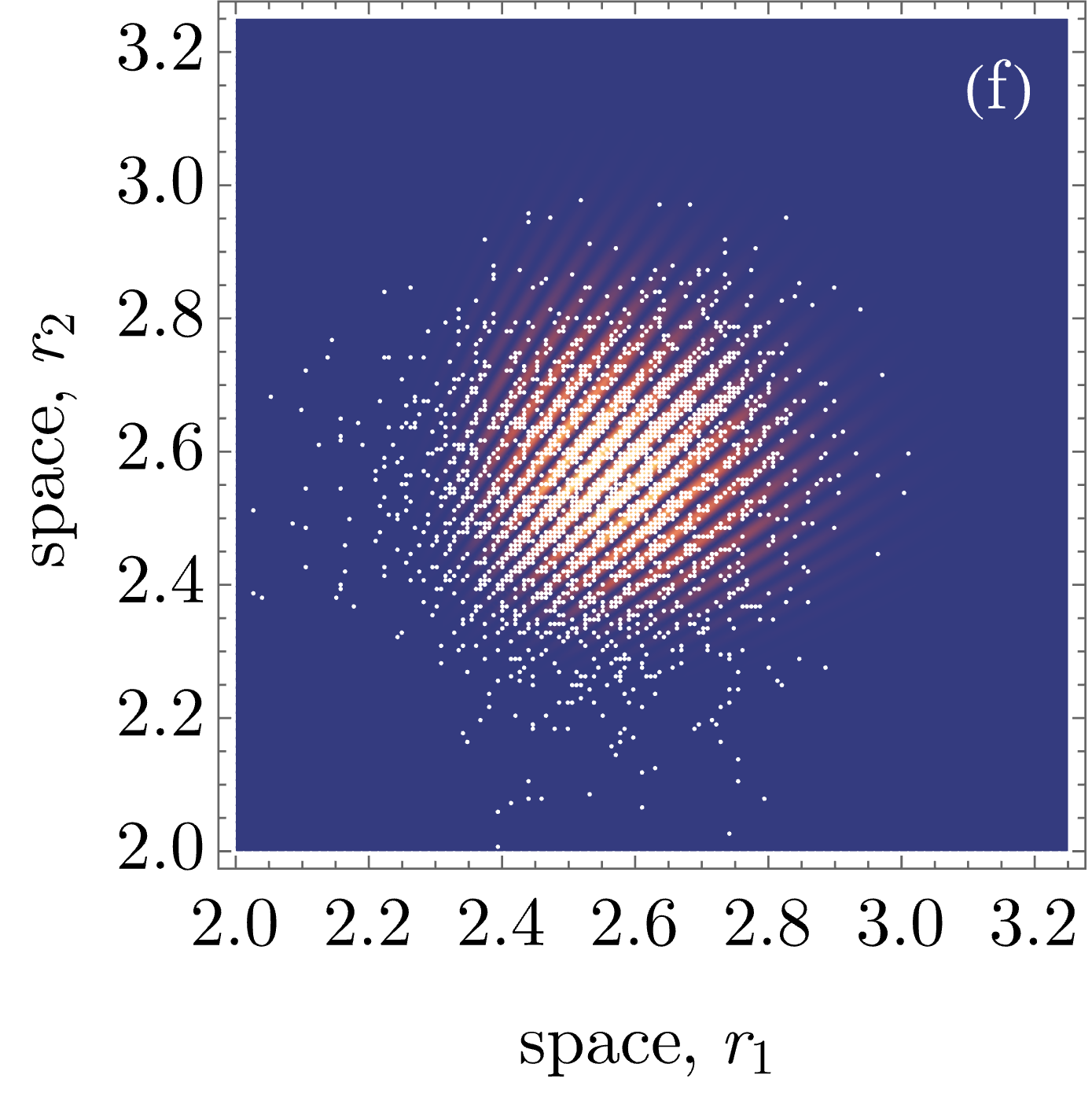}}
 \caption{(a)--(c) Weak value trajectories of a single photon in Schwarzschild spacetime with $k_0 /\sigma = 15$. (a) Balanced superposition of inwardly and outwardly directed wavepackets, (b) $\alpha = 3/4$, (c) $\alpha = 1$. (d)-(f) Two-photon weak value trajectories in Schwarzschild spacetime.(d) Trajectories of both photons on a common $(t,r)$ coordinate system, with initial conditions distributed according to the quantum-mechanical density. (e)-(f) Randomly sampled Gaussian distribution of initial trajectories for both photons at (e) $t = -1$ and (f) $t = 0$ in the $(r_1, r_2)$ plane. Both photons have $k_0/\sigma = 20$. }
 \label{fig:trajectories}
\end{figure*}

As mentioned above, we consider the optical approximation for the wavepacket $f(k)$, treating it as a narrowly focused beam with transverse width significantly narrower than the radial distance from the light source. This allows us to consider the radial solutions as approximately plane waves, neglecting the transverse degrees of freedom of the propagation \cite{peltola2009local,Carlip2007}. By invoking the optical approximation, we also restrict our analysis to the single-particle sector of the field i.e.\ where the wavepacket is peaked far away from the negative frequencies where particle creation effects typically emerge \cite{Gunter2009,Riek2017}. In this approximation, the Klein-Gordon equation becomes 
\begin{align}
    ( \Box + U(r) ) \psi(t, r^\star ) &= 0 
    \vt 
\end{align}
where $\Box \equiv \p^2/\p t^2 - \p^2/\p r^{\star 2}$ is the d'Alembert operator, $U(r) = f'(r)f(r)/r$ is an effective potential that approaches zero at the horizon and asymptotically far from it, and $\psi(t,r^\star)$ is the radial Klein-Gordon wavefunction. For large masses, $U(r)$ can be approximated as a constant, implying dispersive solutions of the form $\psi_\pm(t,r^\star) \propto \exp ( -i k ( t \pm \mu r^\star) )$ where $\mu = \sqrt{1 - U/k^2}$. For wavepackets peaked at optical frequencies, the effective mass term $U/k^2$ can be neglected, leaving pure plane wave solutions (physically this corresponds to the regime where the length scale of the wavepacket is much smaller than that of the mass). The resulting wave equation written in terms of tortoise coordinates adequately describes the effect of the metric geometry on the dynamics. We remark that the Penrose-Chernikov-Tagirov coupling $-R/6$ does not arise here since the Ricci scalar vanishes in the Schwarzschild spacetime. Now in this limit, one can express the Klein-Gordon momentum and Hamiltonian operators in differential form as $\h p = -i \p/\p r^\star, \h H = i \p/\p t$, giving
\begin{align}
    v(t,r) &=f(r) \frac{j^1(t,r^\star)}{j^0(t,r^\star)}
    \vt 
    \label{eq9}
\end{align}
where $j^\mu (t,r^\star) = 2\mrm{Im} \psi^\star(t,r^\star) \p^\mu \psi(t,r^\star)$ are the components of the Klein-Gordon conserved current with $\psi(t,r^\star) \equiv \langle \phi | \psi(t) \rangle$. The velocity field in Eq.\ (\ref{eq9}), constructed from the components of the Klein-Gordon 4-current, is explicitly Lorentz covariant, satisfying the relativistic velocity addition rule. The components $j^\mu$ satisfy a continuity equation in the $(t,r^\star)$ coordinates, $\p_\mu j^\mu  = 0$.

Let us now evaluate the trajectories predicted by our theory. We consider a Gaussian wavepacket in a superposition of inward and outwardly directed radial trajectories, i.e.\ 
$f(k) = \sqrt{\alpha} f_+(k) + \sqrt{1 - \alpha} f_-(k)$ with $0 \leq \alpha \leq 1, f_\pm(k) = (2\pi\sigma)^{-1/4} \exp ( - ( k \mp k_0)^2/4\sigma^2)$, where $k_0, \sigma$ are respectively the central frequency and bandwidth. In the optical approximation, $k_0/\sigma \gg 1$, meaning that the wavepackets have support on $k>0, k<0$ respectively. 
% It can be easily shown that the radial and time components of the conserved current are given by $j^\mu(t,r^\star) = \sqrt{2/\pi}\sigma( \rho_+ + (-1)^\mu \rho_- + J_\mu \sqrt{\rho_+ \rho_-})$ for $\mu = 0, 1$,
% \begin{align}
%     j^1(t,r^\star) &= \sqrt{2/\pi}\sigma ( \rho_\mrm{R} - \rho_\mrm{L} + J_1 \sqrt{\rho_\mrm{R} \rho_\mrm{L}} )
%     \vt  \\
%     j^0(t,r^\star) &= \sqrt{2/\pi} \sigma ( \rho_\mrm{R} + \rho_\mrm{L} + J_0 \sqrt{\rho_\mrm{R} \rho_\mrm{L}} )
%     \vt 
% \end{align}
% where $\rho_+ = \exp( -2U^2\sigma^2), \rho_- = \exp ( - 2 V^2\sigma^2)$ having defined the lightcone coordinates $U = t - r^\star, V = t + r^\star$ and $J_1 = (4\sigma^2t/k_0) \sin(2k_0r^\star), J_0 = \cos(2k_0r^\star) - (2\sigma^2t/k_0)\sin(2k_0r^\star)$. 
% \begin{align}
%     J_1 &= \frac{4\sigma^2t}{k_0} \sin(2k_0r^\star) 
%     \nonvt \\
%     J_0 &= \cos(2k_0r^\star) - \frac{2\sigma^2t}{k_0} \sin(2k_0r^\star)
%     \nonvt 
% \end{align}
We emphasise that our choice of coordinates is merely for convenience, but yields equivalent expressions for the Klein-Gordon current and density as would be obtained if utilising local coordinates corresponding to an observer's proper time and distance. This is because a transformation to such local coordinates is accompanied by a concomitant redshift $\sim \sqrt{f(r)}$ that scales the frequency and bandwidth of the modes, leaving $j^\mu$ invariant.

The expressions for $j^\mu$ can be straightforwardly computed (see Supplemental Materials), which we insert into Eq.\ (\ref{eq9}) and solve $v(t,r) = \D r / \D t$ numerically for a distribution of initial conditions that matches the initial Klein-Gordon density. While the velocity field is formally expressed in terms of the Klein-Gordon current and density in tortoise coordinates, we plot trajectories in Schwarzschild coordinates to capture the curvature effects of the black hole. This yields the trajectories shown in Fig.\ \ref{fig:trajectories}(a)-(c) for variously weighted superpositions. Underlaid in each plot is the probability density as predicted by quantum field theory. We observe how the density of trajectories matches the quantum-mechanical density, as guaranteed by the continuity equation, bunching in regions of constructive interference and anti-bunching in regions of destructive interference. The trajectories in regions of destructive interference have coordinate velocities that exceed the local speed of light. This is consistent with general relativity, which only requires that light travels at $c$ locally, as well as our measurement framework, which gives rise to a velocity field obtained in a particular reference frame.

These results are fully interpretable within the paradigm of quantum theory in curved spacetime. On one hand, the measurement-based velocity field can be understood as an ensemble average obtained through repeated weak measurements by local observers. Similarly, the Klein-Gordon velocity field describes the flow of probability current in the Schwarzschild spacetime, which obeys the usual time-evolution governed by relativistic quantum mechanics. On the other hand, Bohmian mechanics posits that photons follow \textit{actual}, deterministic trajectories in spacetime. Remarkably, we find that such classical trajectories arise naturally as null geodesics in an admissible classical solution to Einstein's field equations. Such a solution must allow for classical worldlines whose Schwarzschild coordinate velocity exceeds the speed of light, as well as admitting trajectories that become stationary (see Fig.\ \ref{fig:trajectories}(a)-(c)). As we show in the Supplemental Materials, a spacetime that fulfils these requirements and can be obtained directly from the Arnowitt-Deser-Misner formalism \cite{ADMPhysRev.116.1322,maclaurin2024falling}, which we dub Schwarzschild-Alcubierre, is described by the following line element:
\begin{align}
    \D s^2 &= - (1 - v_s^2 ) f(r) \D t^2 + f(r)^{-1} \D r^2 - 2 v_s \D r \D t 
    \vt 
    \label{eq12}
\end{align}
where $f(r)$ is the usual Schwarzschild metric function and $v_s$ is a function related to the weak value and Klein-Gordon velocity fields by $v_s = (j^1/j^0 - 1 ) \mrm{sgn}(j^1/j^0)$. For null geodesics, it can be easily shown that the line element Eq.\ (\ref{eq12}) gives the coordinate velocity 
\begin{align}
    \frac{\D r}{\D t} = f(r) ( v_s + \mrm{sgn} ( j^1/ j^0 ) ) \equiv f(r) \frac{j^1(t,r^\star)}{j^0(t,r^\star)} ,
\end{align}
where $\mrm{sgn}(j^1/j^0) = + 1$ for $j^1/j^0 \geq 0$ and $-1$ elsewhere. That is, classical null geodesics in this spacetime are equivalent to the average quantum-mechanical trajectories computed using weak values/Klein-Gordon fields. Equation (\ref{eq12}) features several intuitive limiting cases. When $v_s = 0$, one recovers Schwarzschild spacetime, and photons follow the worldlines given by $\D r /\D t = f(r)$. This corresponds to the case where the quantum-mechanical wavepackets are purely inwardly or outwardly directed. For intermediate values of $\alpha$ (i.e.\ when interference effects are present), Eq.\ (\ref{eq12}) is an interpolation of the Schwarzschild and Alcubierre spacetimes, the latter being a non-globally-hyperbolic spacetime proposed by Alcubierre as a way of engineering a so-called warp drive that allows for superluminal travel \cite{alcubierre1994warp}. Such superluminality is achieved through a local expansion (contraction) in the region behind (in front of) a local observer. In this interpolated regime, the classical photon trajectories experience the combined effects of near-horizon curvature due to the black hole, as well as exotic effects like superluminal (and vanishing) velocities. When $m = 0$, one recovers a special case of the Alcubierre metric, the solution of which was first pointed out in Ref.\ \cite{foo2022relativistic} for the case of photon trajectories in Minkowski spacetime \nocite{lantigua2024hartman}. 

The implication of the above {is} that our Schwarzschild-Alcubierre spacetime and the deterministic photon trajectories it admits is fully compatible with the predictions of quantum theory in curved spacetime. That is, if one adopts the assumptions of Bohmian mechanics, the Schwarzschild-Alcubierre spacetime may be interpreted as a ``guiding metric'' that is a function of the wavefunction itself (in the same spirit as the so-called ``guiding potential'' that arises in nonrelativistic Bohmian mechanics \cite{bohmPhysRev.85.166}) and photons propagate on this background in a manner that is consistent with the average velocity field obtained via weak measurements in the Schwarzschild spacetime. Importantly, since Bohmian mechanics is merely a framework for interpreting the predictions of quantum theory, it is by construction empirically indistinguishable from any other interpretation. Thus, despite the unusual properties of the ``classical'' (insofar as it satisfies the classical field equations) metric Eq.\ (\ref{eq12}), most notably the requirement of exotic matter (e.g.\ negative energy density) to source it \cite{alcubierre1994warp}, everything we have presented is consistent with standard quantum theory.

To illustrate the generality of our approach, let us extend our framework to describe interference between two indistinguishable photons in the Schwarzschild spacetime. As with the single photon case, we are guided by the three pillars of our unifying framework; the weak value ``ensemble average'' approach, the Klein-Gordon dynamical approach, and the connection to classical trajectories via a general relativistic metric. Inspired by the setup of Ref.\ \cite{fooPhysRevA.109.022229}, we consider two initially separable and indistinguishable photons with wavevectors directed towards each other, and nondirectional detectors that are agnostic to the identity of the respective photons. We are interested in the scenario where each detector registers a single detection event, and assume they are operative on the same timeslice in tortoise coordinates. Let us propose the following expressions for the weak-value velocities of the respective photons:
\begin{align}
    v_1(r_1, r_2, t) &= f(r_1) \frac{\langle \hat{\mathbf{p}}_{Aw} \rangle}{\langle \hat{\mathbf{H}}_{Aw} \rangle} 
    \vt 
    \label{eq14}
    \\
    v_2(r_1, r_2, t) &= f(r_2) \frac{\langle \hat{\mathbf{p}}_{Bw} \rangle}{\langle \hat{\mathbf{H}}_{Bw} \rangle} 
    \vt 
    \label{eq15}
\end{align}
where $\hat{\mathbf{p}}_D, \hat{\mathbf{H}}_D$ are momentum and Hamiltonian operators associated with detectors $D = A, B$ (as discussed further in the Supplemental Materials), symmetrised appropriately to encode the agnosticism of the detectors to the identity of the photons.
% \begin{align} 
%     \hat{\mathbf{m}}_A &= \hat{m}_A \hat{\Pi}_{1_A} \otimes \hat{I}_B \hat{\Pi}_{2_B} + \hat{m}_A \hat{\Pi}_{2_A} \otimes \hat{I}_B \hat{\Pi}_{1_B}
% \end{align}
% for $m = p, H$ and $\hat{\Pi}_{i_D}$ are momentum projectors acting on the Hilbert space of particle $i = 1, 2$ and measured by detector $D = A, B$. The form of $\hat{\mathbf{p}}_B, \hat{\mathbf{H}}_B$ follow analogously with the replacement $(A \leftrightarrow B)$. 
The postselection is modelled as a conditional measurement on the position of both photons, that is a superposition of detector $A$ measuring the photon in mode 1 and detector $B$ measuring the photon in mode 2, and vice versa. Our main result for the two-photon case (as we show in detail in the Supplemental Materials), is that the same velocity field can be reproduced using an explicitly covariant multiparticle (scalar) Klein-Gordon theory for an initial position-and-time symmetrised state of the two photons.

In Fig.\ \ref{fig:trajectories}(d)-(f), we have plotted the resulting trajectories of the two photons. In Fig.\ \ref{fig:trajectories}(d), we represent these trajectories on a common $(t,r)$ coordinate system. Starting in a separable state, the photons begin to interfere as they approach each other due to the path-entanglement that is generated through the correlated measurements performed by the agnostic detectors. Figs.\ \ref{fig:trajectories}(e)-(f) show a randomly sampled distribution of initial conditions for the trajectories of both photons, centred on the joint probability distribution predicted by standard quantum theory. The symmetry of the density reflects the indistinguishability of the photons. At the earlier time, the distribution of trajectories maintains a (redshifted) Gaussian distribution, while near $t \sim 0$, the photons exhibit interference due to the above mentioned correlations between them, which is a manifestation of nonclassical photon bunching. 

As in the single-photon scenario, we can draw a connection between the operationally-derived average trajectories with classical geodesics in a general relativistic spacetime. The natural generalisation of Eq.\ (\ref{eq12}) is, 
\begin{align}
    \D s^2 &= - ( 1 - \mathbf{v}_s^2 ) f(r_i) \D t^2 + f(r_i)^{-1} \D r^2 - 2 \mathbf{v}_s \D r \D t 
    \label{eq20}
\end{align}
where $\mathbf{v}_s = ( | v_i(\mathbf{R}_1, \mathbf{R}_2) | - 1 ) \mrm{sgn} ( v_i ( \mathbf{R}_1, \mathbf{R}_2 ) )$. Note in particular that Eq.\ (\ref{eq20}) should be understood as the metric determined locally by an observer i.e.\ one who can perform local measurements of energy and momentum of the photons. The subscript $i = 1,2$ thus denotes the local metric felt by either photon. It can be straightforwardly shown that this choice of $\mathbf{v}_s$ gives rise to the correct quantum-mechanical velocities, Eq.\ (\ref{eq14}) and (\ref{eq15}). The most notable feature of Eq.\ (\ref{eq20}) is that it depends not only on local measurements of the photon of interest, but also on the spacetime coordinate of the other photon. This spacetime nonlocality is expected within the Bohmian interpretation of quantum theory, however its meaning within general relativity, which is a local theory, remains an open question. As we have emphasised, the weak value velocities are perfectly compatible with the predictions of curved spacetime quantum field theory. While the metrics in Eq.\ (\ref{eq12}) and (\ref{eq20}) are valid solutions to the classical field equations of general relativity, we have not constructed them from first principles, i.e.\ starting with a stress-energy tensor corresponding to a given matter distribution and solving the resulting classical field equations. The fact that they still give rise to trajectories equivalent to ones obtained via an explicitly quantum-dynamical approach (i.e.\ through post-selected weak measurements of quantum operators) raises the interesting question of whether such metrics may give guidance on problems at the interface of quantum mechanics and general relativity \cite{fooPhysRevLett.129.181301,fooPhysRevD.107.045014,giacomini10.1116/5.0070018,ruizdoi:10.1073/pnas.1616427114,Zych2019,delaHamette2023}.

In conclusion, we have presented a unifying framework for interpreting quantum-mechanical velocities of relativistic particles in curved spacetime as undergoing deterministic dynamics. While we have specifically studied optical frequency photons propagating in a Schwarzschild spacetime, our framework is general and should be applicable to other scenarios. Moreover, it is operationally motivated, insofar as the weak value prescription for obtaining the velocity field can in-principle be implemented experimentally through appropriate weak measurements of momentum and energy. While such an experiment would no doubt be challenging, space-based tests of curvature effects on photon interference may be within reach of current experiments \cite{panPhysRevLett.133.020201}. On the other hand, these average trajectories may be interpreted as the worldlines of \textit{classical} photons propagating in a hypothetical hybrid spacetime whose curvature is a function of the quantum wavefunction, within which such worldlines are simply geodesics.

\bibliography{main} 

\clearpage

\title{Supplemental Material: General relativistic particle trajectories via quantum mechanical weak values and the Schwarzschild-Alcubierre spacetime}

\maketitle

\setcounter{page}{1}
\renewcommand{\theequation}{S\arabic{equation}}
\setcounter{equation}{0}

\begin{widetext}

\subsection{Equivalence of the Klein-Gordon Velocity Field from Weak Values}

\noindent In this section we derive the equivalence between the weak value velocity and the Klein-Gordon velocity in Schwarzschild spacetime. Recall from the main text,
\begin{align}
    v(t,r) &= f(r) \frac{\mrm{Re} \langle \phi | \hat p | \psi(t) \rangle}{\mrm{Re} \langle \phi | \hat H | \psi(t) \rangle} 
\end{align} 
Anticipating our invocation of the optical approximation, which allows us to consider approximate radial plane wave solutions, we focus on particle dynamics in the radial direction only. We consider postselected position eigenstates defined with respect to tortoise coordinates, leading to,
\begin{align}
    v(t,r) &= f(r) \frac{\mathrm{Re}\langle r^* | \hat{p} | \psi(t) \rangle}{\mathrm{Re}\langle r^* | \hat{H} | \psi(t) \rangle}
\end{align}
To draw a connection with the Klein-Gordon velocity, we utilise the differential forms of the Klein-Gordon momentum and Hamiltonian operators. The momentum operator $\hat{p}$ generates spatial displacements, while $\h H$ generates time-evolution. Since we are dealing with radial plane waves, these take the simple form:
\begin{align}
    \langle r^* | \hat{p} | \psi(t) \rangle &= -i\partial_{r^*}\psi(t,r^*) = -i\partial^{r^*} \psi(t,r^*)
    \vt 
\\ 
    \langle r^* | \hat{H} | \psi(t) \rangle &= i\partial_{t}\psi(t,r^*) = -i\partial^{t} \psi(t,r^*)
    \vt 
\end{align}
The weak-value expressions can be evaluated using these identifications for momentum and energy,

\begin{align}
    \langle r^* | \hat{p}_w | \psi(t) \rangle &= \mrm{Re} \frac{\langle r^* | \hat{p} | \psi(t) \rangle}{\langle r^* | \psi(t) \rangle} 
    % \nonumber 
    % \\
    % &= \mathrm{Re} \frac{\langle \psi(t) | r^* \rangle \langle r^* | \hat{k} | \psi(t) \rangle}{\langle \psi(t) | \psi(t) \rangle} \nonumber \\
    = \mathrm{Re} \frac{ (-i) \psi^*(t,r^*) \partial^{r^*} \psi(t,r^*)}{|\psi(t,r^*)|^2}
\vt 
\\
    \langle r^* | \hat{H}_w | \psi(t) \rangle &= \mathrm{Re} \frac{\langle r^* | \hat{H} | \psi(t) \rangle}{\langle r^* | \psi(t) \rangle} 
    % \nonumber \\
    % &= \mathrm{Re} \frac{\langle \psi(t) | r^* \rangle \langle     r^* | \hat{H} | \psi(t) \rangle}{\langle \psi(t) | \psi(t) \rangle} \nonumber \\
    % &
    = \mathrm{Re} \frac{(-i) \psi^*(t,r^*) \partial^{t} \psi(t,r^*)}{|\psi(t,r^*)|^2}
\end{align}
Dividing the two weak values yields,
\begin{align}
v(t,r^*) &= f(r) \frac{\mathrm{Re}(-i) \psi^*(t,r^*) \partial^{r^*} \psi(t,r^*)}{\mathrm{Re} (-i) \psi^*(t,r^*) \partial^{t} \psi(t,r^*)} \nonumber \\
&= f(r) \frac{2 \mathrm{Im} \psi^*(t,r^*) \partial^{r^*} \psi(t,r^*)}{2 \mathrm{Im} \psi^*(t,r^*) \partial^{t} \psi(t,r^*)}
\end{align}
The numerator and denominator are the Klein-Gordon conserved probability current and density respectively.

\subsection{Klein-Gordon Equation in the Optical Approximation}

\noindent In this section we describe the steps and approximations used to reduce the solutions to the Klein-Gordon equation to radial plane waves. Beginning with the Klein-Gordon equation in an arbitrary curved spacetime, $g^{\mu\nu} D_\mu D_\nu \psi = 0$, we then invoke the optical approximation for which we obtain the simplified equation,
\begin{align}
    ( \Box + U(r) ) \psi(t, r^\star ) &= 0 
    \vt 
    \label{eq19app}
\end{align}
where $\Box \equiv \p^2/\p t^2 - \p^2/\p r^{\star 2}$ is the d'Alembert operator, $U(r) = f'(r)f(r)/r$ is an effective potential, and $\psi(t,r^\star)$ is the radial Klein-Gordon wavefunction. For large mass black holes, $U(r)$ may be approximated as a constant due to its slow decay. To obtain the solution to Eq.\ (\ref{eq19app}), we make the ansatz:
\begin{align}
    \psi(t, r^\star ) &= C(r^*) e^{-ik t}
    \vt 
\end{align}
leading to a separation of variables and the following ordinary differential equation for $C(r^*)$:
\begin{equation}
    C''(r^*) + (k^2 - U(r)) C(r^*) = 0.
\end{equation}
The general solution for $C(r^*)$ in regions where $U(r) < k^2$ is:
\begin{align}
C(r^*) &= \alpha \exp \Big[ i \sqrt{k^2 - U}r^\star \Big] + \beta \exp\Big[ -i \sqrt{k^2 - U}r^\star \Big]
\vt 
\end{align}
Accordingly, the general solution for the radial wave function $\psi(t, r^*)$ is given by
\begin{align}
    \psi(t, r^\star ) &= \alpha \exp \Big[ - i k \Big(t - \sqrt{1 - U(r)/k^2}r^* \Big) \Big] + \beta \exp\Big[ - i k \Big(t + \sqrt{1 - U(r)/k^2} r^* \Big) \Big] 
    \vt 
\end{align}
Considering a redshift factor $\sqrt{1-2m/r}$ due to the distant observer and noting that for large $k$, due to the large optical frequencies we are considering, the quantity $U(r)/k^2 = (2m)/(r^3k^2)$ is negligible. In this limit the solutions thus reduce to pure plane waves,
\begin{align}
    \psi(t, r^\star ) &= \alpha \exp \Big[ i k \left(t + r^*\right) \Big] + \beta \exp \Big[ i k \left(t - r^*\right) \Big] 
    \vt 
\end{align}
as desired. 

%\begin{align}
%    \left( \p^2/\p t^2 - \p^2/\p r^{\star 2} + U(r) \right) C(r^*) e^{ik t} &= 0 \\
%    -C(r^*) k^2 e^{ik t} - C^{''}(r^*) e^{i k t} + U(r)C(r^*) e^{ik t} &= 0 \\
%    C^{''}(r^*) &= -C(r^*) (k^2 - U(r)) \\
%    C(r^*) &= \alpha \sin{\sqrt{k^2 -U(r)} r^*} + \beta \cos{\sqrt{k^2 -U(r)} r^*} \\
%    \text{If the optical limit, where $U(r) \textless k^2$}: \\
%    C(r^*) &= \alpha e^{i \sqrt{k^2 - U}r^*} + \beta e^{-i \sqrt{k^2 - U}r^*}
%    \vt 
%\end{align}

%And for $\psi(t,r^*):$
%\begin{align}
%    \psi(t, r^\star ) &= \alpha e^{i k \left(t + \sqrt{1 - \frac{u(r)}{k^2}}r^*\right)} + \beta e^{i k \left(t - \sqrt{1 - \frac{u(r)}{k^2}}r^*\right)}
%    \vt 
%\end{align}

\subsection{Radial and Time Components of the Conserved Current}

\noindent Here, we compute the explicit expressions for the radial and time components of the Klein-Gordon conserved current. We consider a Gaussian wavepacket initially configured in a superposition of inwardly and outwardly directed radial trajectories in Schwarzschild spacetime represented in tortoise coordinates, $r^*$. The wavepacket profile in momentum space is constructed as a linear combination:

\begin{align}
f(k) = \sqrt{\alpha} f_+(k) + \sqrt{1 - \alpha} f_-(k) 
\end{align}

\noindent where $0 \leq \alpha \leq 1$ and

\begin{align}
f_\pm(k) &= \left( \frac{1}{2\pi \sigma^2} \right)^{1/4} \exp \Big[ - \frac{(k\mp k_0)^2}{4\sigma^2} \Big] 
\end{align}

\noindent where $k_0,\sigma$ are the central frequency and bandwidth respectively. The corresponding wavefunction in position space is:

\begin{align}
\psi(t,r^*) = \sqrt{\alpha} \int_{-\infty}^{+\infty} \D k \: f_+ (k)e^{-i|k|t+ikr^*} + \sqrt{1 - \alpha} \int_{-\infty}^{+\infty} \D k \: f_-(k)e^{-i|k|t+ikr^*}.
\end{align}

\noindent In the quantum optical regime, $k_0/\sigma \gg 1$, i.e.\ each wavepacket is centred far from zero frequencies. Thus, each wavepacket component $f_\pm(k)$ only has support on the positive or negative frequencies respectively. This restriction affords a clear delineation of inward- and outward-moving components, thereby simplifying subsequent calculations of the probability current and density. We can therefore write,

\begin{align}
    \psi(t, r^*) &\simeq \sqrt{\alpha} \int_{0}^{+\infty} \D k f_{+}(k)e^{-i|k|t+ikr^*}  + \sqrt{1 - \alpha} \int_{-\infty}^{0} \D k f_{-}(k)e^{-i|k|t+ikr^*}
\\
    &= \sqrt{\alpha} \int_{0}^{+\infty} \D k f_{+}(k)e^{-ik(t-  r^*)} + \sqrt{1 - \alpha} \int_{0}^{+\infty} \D k f_{-}(-k)e^{-ik(t+r^*)}
\\
    \label{Seq30}
    &\simeq \sqrt{\alpha} \int_{-\infty}^{+\infty} \D k f_{+}(k)e^{-ik(t-r^*)} + \sqrt{1 - \alpha} \int_{-\infty}^{+\infty} \D k f_{-}(-k)e^{-ik(t+r^*)} 
\end{align}
In the second line, we have eliminated the absolute value signs due to the integrals' strictly positive or negative domains, and then re-extended the integral bounds under the given assumptions on the wavepacket localisation. Performing the integrals gives simply,

\begin{align}\label{Seq31}
    \psi(t, r^*) &= \left( \frac{2\sigma^2}{\pi} \right)^{1/4} \Bigg[ \sqrt{\alpha } \exp \Big( - (t - r^*) (i k_0 + (t - r^*) \sigma^2) \Big) + \sqrt{1 - \alpha} \exp \Big( - (t + r^*) (i k_0 + (t + r^*) \sigma^2) \Big) \Bigg] 
\end{align}

%where \( \mathcal{J}_0 = (2\sigma^2/\pi)^{1/4} \) and
%\begin{align}
%V_{0+} &= i k_0 + (t - r^*)^2 \sigma^2 \\
%V_{0-} &= i k_0 + (t + r^*)^2 \sigma^2 
%\end{align}

Inserting Eq. \ref{Seq31} into the expressions for the relativistic Klein-Gordon probability current and density,

\begin{align}
    j^1(t, r^*) &= 2\mrm{Im} \Big[ \psi^*(t, r^*) \partial^{t} \psi(t, r^*) \Big] 
\vt 
\\
    j^0(t, r^*) &= 2\mrm{Im} \Big[ \psi^*(t, r^*) \partial^{r^*} \psi(t, r^*) \Big] 
\end{align}

gives the radial and time components of the conserved current: 

%\begin{equation}
%\begin{aligned}
%&j_0(t,r^*) = \frac{1}{2} \sqrt{\frac{2}{\pi}} \sigma \\
%&\left( e^{-2(t-r^*)^2\sigma^2} - e^{-2(t+r^*)^2\sigma^2} + S_0 %e^{-2(t^2+{r^*}^2)\sigma^2} \right)
%\end{aligned}
%\end{equation}

%\begin{equation}
%\begin{aligned}
%&j_1(t,r^*) = \frac{1}{2} \sqrt{\frac{2}{\pi}} \sigma \\
%&\left( e^{-2(t-r^*)^2\sigma^2} + e^{-2(t+r^*)^2\sigma^2} + T_0 %e^{-2(t^2+{r^*}^2)\sigma^2} \right)
%\end{aligned}
%\end{equation}

%Where:
%\begin{equation}
%S_0 = 4 \frac{\sigma^2 t}{k_0} \sin{(2k_0r^*)}
%\end{equation}
%\begin{equation}
%T_0 = \cos{(2k_0 r^*)} - 2\frac{\sigma^2 t}{k_0} \sin{(2k_0 r^*)}
%\end{equation}

%$j^\mu(t,r^\star) = \sqrt{2/\pi}\sigma( \rho_+ + (-1)^\mu \rho_- + J_\mu \sqrt{\rho_+ \rho_-})$ for $\mu = 0, 1$,

\begin{align}     
    j^1(t,r^\star) &= \sqrt{\frac{2}{\pi}} \sigma \Big[ \rho_+ - \rho_- + J_1 \sqrt{\rho_+ \rho_-} \Big]
    \vt  \\
    j^0(t,r^\star) &= \sqrt{\frac{2}{\pi}} \sigma \Big[ \rho_+ + \rho_- + J_0 \sqrt{\rho_+ \rho_-} \Big] 
\vt 
\end{align}

where $\rho_\pm = \exp( -2(t\mp r^\star)^2\sigma^2)$ and 
\begin{align}
     J_1 &= \frac{4\sigma^2t}{k_0} \sin(2k_0r^\star) 
     \vt \\
     J_0 &= \cos(2k_0r^\star) - \frac{2\sigma^2t}{k_0} \sin(2k_0r^\star)
     \vt  
\end{align}

\subsection{Derivation of the Schwarzschild-Alcubierre Spacetime Metric}

In this section we make use of the framework recently introduced in Ref.\ \cite{maclaurin2024falling}, which utilises the Arnowitt-Deser-Misner (ADM) formalism for general relativity to derive the Schwarzschild-Alcubierre spacetime metric.

Following the generalised warp field construction from Ref.\ \cite{maclaurin2024falling}, we introduce modifications to the Schwarzschild metric to encompass a ``warp bubble.'' This is achieved by introducing a warp shift vector field $b$ and a warp lapse $\alpha$, which together modify the Schwarzschild interval. The warp shift is specifically orthogonal to the four-velocity of the static observers $u$ in the original Schwarzschild metric, ensuring that the new metric maintains Lorentzian properties. The interpolated metric tensor is formally expressed as,
\begin{align}\label{sEq1}
g_{\mu\nu}^{\mathrm{warp}} &:= (1 - \alpha^2 + \langle b, b \rangle)u^{\mu}u^{\nu} - u^{\mu}b^{\nu} - b^{\mu}u^{\nu} + g_{\mu\nu}^{\mathrm{original}}
\end{align}
where $g_{\mu\nu}^{\mathrm{original}} \equiv g_{\mu\nu}^\mathrm{Sch.}$ is the Schwarzschild metric:

\begin{align}\label{sEq2}
g_{\mu\nu}^{\mathrm{Sch.}} &= \begin{pmatrix} -f(r) & 0 \\ 
0 & f(r)^{-1}
\end{pmatrix}
\end{align}
with $f(r) = (1 - 2m/r)$. The modified warp metric components are:
\begin{align}
\label{sEq3}
    g_{00}^{\mathrm{warp}} &= (1 - \alpha^2 + \langle b, b \rangle) u_0^2 -f(r) - 2 u_0 b_0 
    \vt 
    \\
    g_{01}^{\mathrm{warp}} &= (1 - \alpha^2 + \langle b, b \rangle) u_0 u_1 - u_0 b_1 - u_1 b_0 
    \vt
    \\
    g_{10}^{\mathrm{warp}} &= (1 - \alpha^2 + \langle b, b \rangle) u_1 u_0 - u_1 b_0 - u_0 b_1 
    \vt 
    \\
    g_{11}^{\mathrm{warp}} &= (1 - \alpha^2 + \langle b, b \rangle) u_1^2 - 2 u_1 b_1 + f(r)^{-1} 
    \vt 
\end{align}
For static observers defined by $u_{\mu} = (-1, 0, 0, 0)$, warp shift $b_{\mu} = (0, -v_s, 0, 0)$, and warp lapse $\alpha = 1$, the scalar product within the Schwarzschild background yields $\langle b, b \rangle = f(r) v_s^2$. The resulting modified warp metric components become:
\begin{align}
    g_{00}^{\mathrm{warp}} &= -(1 - v_s^2)f(r), 
\vt 
\\
    g_{01}^{\mathrm{warp}} &= g_{10}^{\mathrm{warp}} = -v_s, 
\vt 
\\
    g_{11}^{\mathrm{warp}} &= f(r)^{-1}
\vt 
\end{align}

\noindent These components form the Schwarzschild-Alcubierre metric,

\begin{align}\label{sEq4}
\D s^2 &= -(1 - v_s^2) f(r) \D t^2 + f(r)^{-1} \D r^2 - 2v_s \D r \D t
\end{align}

\noindent The corresponding null geodesic condition is given by:
\begin{align}\label{sEq5}
0 &= -(1 - v_s^2) f(r) + f(r)^{-1} \frac{dr^2}{dt^2} - 2v_s \frac{dr}{dt} 
\end{align}

\noindent Solving Eq.\ (\ref{sEq5}) for $\D r /\D t$, utilising Eq.\ref{eq9}, we find:
\begin{align}
    \frac{\D r}{\D t} = f(r) ( v_s + \mrm{sgn} ( j^1/ j^0 ) ) \equiv f(r) \frac{j^1(t,r^\star)}{j^0(t,r^\star)} 
\end{align}
as shown in the main text.

\subsection{Equivalence of Weak Value and QFT Velocities for Two Photons}

\noindent In this section, we demonstrate the equivalence between the weak value and QFT approaches for obtaining the average velocity fields (equally, the flow fields for probability current) of two indistinguishable photons. We consider an initially separable state of the two photons:
\begin{align}\label{eq5.12}
    | \psi(t) \rangle &= \hat{a}_{f_1}^{\dagger} \hat{a}_{f_2}^{\dagger} | 0 \rangle | 0 \rangle = | f(k_1) \rangle | f(k_2) \rangle 
\end{align}
where 
\begin{align}\label{eq2}
    \hat{a}_{f_i}^{\dagger} | 0 \rangle &= \int\D k_i \: e^{-iE (k_i)t} f(k_i) \hat{a}_{k_i}^{\dagger} | 0 \rangle 
\end{align}
Here $\hat{a}_{k_i}^\dagger$ is the creation operator that creates a photon in the plane wave mode $k_i$, i.e.\ $\hat{a}_{k_i}^\dagger | 0 \rangle \equiv | k_i \rangle$. $E(k_i) = |k|$ is the energy of the photon in mode $k_i$ while we assume $f_i(k_i)$ to be Gaussian wavepackets of the form,
\begin{align}
    f_i(k_i) &= \left( \frac{1}{2\pi\sigma^2} \right)^2 \exp \left[ - \frac{(k-k_0)^2}{4\sigma^2} \right] 
\end{align}
where $k_0$ is the centre frequency and $\sigma$ the bandwidth. in the optical approximation, such a plane wave description is a good description of a transverse Gaussian mode close to its beam waist. If $|k_0|$ and $\sigma$ are the same for the two photons, they will be indistinguishable apart from their propagation direction. 

As discussed in the main text, we propose the following definitions for the velocity fields of the two photons constructed from weak measurements of their momenta and energy locally at detectors $A$ and $B$:
\begin{align}
    v_1(r_1, r_2, t) &= f(r_1) \frac{\langle \hat{\mathbf{p}}_{Aw} \rangle}{\langle \hat{\mathbf{H}}_{Aw} \rangle} 
    \vt 
    \label{eq5.15}
\\
    v_2(r_1, r_2, t) &= f(r_2) \frac{\langle \hat{\mathbf{p}}_{Bw} \rangle}{\langle \hat{\mathbf{H}}_{Bw} \rangle} 
    \vt 
    \label{eq5.16}
\end{align}
where 
\begin{align} 
    \hat{\mathbf{m}}_A &= \hat{m}_A \hat{\Pi}_{1_A} \otimes \hat{I}_B \hat{\Pi}_{2_B} + \hat{m}_A \hat{\Pi}_{2_A} \otimes \hat{I}_B \hat{\Pi}_{1_B}
    \label{eq56}
\end{align}
with $\hat m = \hat p , \hat H$ $( \hat{\mathbf{m}} = \hat{\mathbf{p}} , \hat{\mathbf{H}})$ representing symmetrised momentum and Hamiltonian operators and $\hat\Pi_{i_D} = | k_{i_D} \rangle\langle k_{i_D}|$ rank-1 projectors identified with the subspace of detector $D = A, B$ sensitive to a photon with directionality $j = 1, 2$ respectively. The measurement operators in Eq.\ (\ref{eq56}) are constructed from weak measurements performed by detector $A$, given a coincident strong value measurement at detector $B$, and vice versa. $\hat{k}_D$, $\hat{H}_D$, and $\hat{I}_D$ are the momentum, Hamiltonian, and identity operators with support on the Hilbert space of detector $D = A, B$. The form of $\hat{\mathbf{p}}_B, \hat{\mathbf{H}}_B$ follows analogously with the replacement $(A \leftrightarrow B)$. As mentioned in the main text, we assume the detectors are non-directional, which will allow interference effects between them to appear. Likewise, we assume that the strong measurements only reveal information about the positions of the photons, and we are interested in the case where each detector registers a single detection event, and focus on the case where the photons have initially oppositely directed wavevectors (i.e.\ one inward- and one outward-directed photon). 

\vspace{10pt}

We consider postselection onto the state $| \bar{r} \rangle$, which models a joint strong measurement of position enacted by both detectors. From field-theoretic considerations, we are motivated to consider both detectors as operating on a single timeslice in a given reference frame (e.g.\ that determined by the original Schwarzschild coordinates), though we are not constrained by this choice--see Ref.\ \cite{fooPhysRevA.109.022229} for further details on the implications of, for example, a timelike arrangement of the detectors. Over repeated runs of the experiment, the detectors gather statistics from the weak measurements of momentum and energy of the incident photons, mapping out the trajectories through successive timeslices. Note that either detector does not exclusively detect one of the photons (or the other). Prior to the interference region, the photons are in a sense distinguishable via the sign of their respective wavevectors, however there is no unique label attached to either. However in the interference region, the detectors cannot identify which photon was measured. In order to interpolate a unique trajectory associated to the respective photons, one needs to interpolate these from the measurement outcomes of both detectors--performed on incrementally successive timeslices--at the end of the proposed experiment.

Let us return to Eq.\ (\ref{eq5.15}) and (\ref{eq5.16}). The weak values are explicitly given by 
\begin{align}\label{eq5.17}
    \prescript{}{\langle \bar{r} | }{\langle \hat{\mathbf{p}}_{Dw} \rangle}_{|\psi(t)\rangle} &= \mathrm{Re} \frac{\langle \bar{r} | \hat{\mathbf{p}}_D | \psi(t)\rangle}{\langle \bar{r} | \psi(t) \rangle}  
    \\
    \label{eq5.18}
    \prescript{}{\langle \bar{r} | }{\langle \hat{\mathbf{H}}_{Dw} \rangle}_{|\psi(t)\rangle} &= \mathrm{Re} \frac{\langle \bar{r} | \hat{\mathbf{H}}_D | \psi(t) \rangle}{\langle \bar{r} | \psi(t) \rangle} 
\end{align}
The state $| \bar{r} \rangle$ takes the explicit form, 
\begin{align}
    | \bar{r} \rangle &= \frac{1}{\sqrt{2}} ( | r_1^\star , 1_A \rangle | r_2^\star , 2_B \rangle + | r_1^\star , 2_A \rangle | r_2^\star , 1_B \rangle ) 
    \vt 
    \label{eq59}
\end{align}
where 
\begin{align}
    | r^\star_i, j_D \rangle &= \int\D k_{j_D} e^{-ik_{j_D}r_i^\star} |k_{j_D} \rangle
    \label{eq60}
\end{align}
and $r_i^\star$ is the tortoise coordinate associated with photon $i$. Notice that though we use $k_{j_D}$ as an integration variable, we retain the notation $j_D$ on the left-hand side of Eq.\ (\ref{eq60}) to label states in the detector subspace (i.e.\ directional momentum eigenstates), which are needed to compute the conditional weak value amplitudes above. The postselected final state in Eq.\ (\ref{eq59}) is constructed from superpositions of position eigenstates (in tortoise coordinates) with support on $k < 0 , k > 0 $ respectively (i.e.\ corresponding to detections of inward- and outward-directed photons). After taking the direct product, one postselects onto detector coincidences such that detector $A$ measures the photon in the ingoing mode and detector $B$ measures the photon in the outgoing one, or vice versa. The factor of $1/\sqrt{2}$ arises after renormalising this conditional state. The same logic here applies to the construction of the symmetrised momentum and Hamiltonian operators in Eq.\ (\ref{eq5.17}) and (\ref{eq5.18}). That is, the detectors measure ``a momentum'' and ``an energy'' at the spacetime location they are situated at, however they remain agnostic to the identity of the detected photon. 

Using these ingredients, we find that the position-space wavefunction $\psi_\mathrm{M} \equiv \psi_\mathrm{M}(t,x_1,x_2) = \langle \bar{r} | \psi(t) \rangle$ is given by:
\begin{align}
    \psi_\mathrm{M} \equiv \psi_\mathrm{M}(t,r_1^\star,r_2^\star) = \langle \bar{r} | \psi (t) \rangle
    &= \frac{1}{\sqrt{2}} ( \psi_1(t,r_1^\star) \psi_2(t,r_2^\star) 
    + \psi_1(t,r_2^\star) \psi_2(t,r_1^\star) )
    \vphantom{\sqrt{\frac{2}{\pi}}}
    \label{eq61}
\end{align}
where
\begin{align}
    \psi_1(t,r_i^\star) &= \int\D k \: e^{-ik (t- r_i^\star)} f(k;k_0) 
    \\
    \psi_2(t,r_i^\star) &= \int\D k \: e^{-ik(t+r_i^\star)} f(k;k_0) 
\end{align}
Here, $\psi_\mathrm{M}(t,r_1^\star,r_2^\star) = \psi_\mathrm{M}(t,r_2^\star,r_1^\star)$ is symmetric under an exchange of the photon positions. Meanwhile, the numerators for the respective weak value expressions are 
\begin{align}\label{eq5.31}
    \langle \bar{r} | \hat{\mathbf{p}}_A | \psi(t) \rangle 
    &= \frac{1}{\sqrt{2}} \left( \psi_{1k} (t,r^\star_1) \psi_2(t,r^\star_2)  - \psi_1(t,r^\star_2) \psi_{2k}(t,r^\star_1) \right) 
    \\
    \langle \bar{r} | \hat{\mathbf{p}}_B | \psi(t) \rangle 
    &= \frac{1}{\sqrt{2}} \left( \psi_{1k}(t,r_2^\star) \psi_2(t,r^\star_1) - \psi_1(t,r^\star_1) \psi_{2k}(t,r_2^\star) \right) 
\end{align}
and the denominators are
\begin{align} 
    \langle \bar{r} | \hat{\mathbf{H}}_A | \psi(t) \rangle 
    &= \frac{1}{\sqrt{2}} \left( \psi_{1k}(t,r^\star_1) \psi_2(t,r^\star_2)  + \psi_1(t,r^\star_2) \psi_{2k}(t,r^\star_1) \right) 
    \\
    \label{eq5.34}
    \langle \bar{r} | \hat{\mathbf{H}}_B | \psi(t) \rangle 
    &= \frac{1}{\sqrt{2}} \left( \psi_1(t,r^\star_1) \psi_{2k}(t,r^\star_2)  + \psi_{1k}(t,r^\star_2) \psi_2(t,r^\star_1) \right) 
\end{align}
having additionally defined
\begin{align}
    \psi_{1k}(t,r_i^\star) &= \int\D k \: e^{-ik(t-r_i^\star)}k f(k;k_0) 
    \\
    \psi_{2k}(t,r_i^\star) &= \int\D k \: e^{-ik(t+r_i^\star)}k f(-k;k_0) 
    \label{eq68}
\end{align}
The weak value velocity fields can be obtained by inserting Eqns.\ (\ref{eq61})-(\ref{eq68}) into the expressions for Eqns.\ (\ref{eq5.17}) and (\ref{eq5.18}). 

Now, we wish to demonstrate the equivalence with the weak value approach for obtaining the average velocity fields with that obtained using an explicitly Lorentz-covariant dynamical theory. Following the multitime Klein-Gordon theory for multiple scalar particles presented in Ref.\ \cite{nikolic2019relativistic}, we first consider the position-space representations of a outward- and inward-going single-particle Klein-Gordon wavefunction, for a Gaussian wavepacket of momenta $f(k;k_0)$, 
\begin{align}
    \psi_1(\mtx_j) &= \int\D k \: e^{-ik(t_j - r_j^\star)} f(k;k_0) 
    \\
    \psi_2(\mtx_j) &= \int\D k \: e^{-ik(t_j + r_j^\star)} f(k;k_0) 
\end{align}
where $\psi_1$, $\psi_2$ are associated with the respective photons. Each photon is identified with a unique spacetime point, which entails the spacetime vector $\mtx_j = (t_j, r_j^\star)$. This identification preserves the Lorentz covariance of the dynamics, in particular allowing for individual Lorentz transformations to be well-defined for the respective photons. As with the measurement-based approach, we consider indistinguishable photons described by the position- and time-symmetrised wavefunction, $\psi_\mathrm{KG}\equiv  \psi_\mathrm{KG}(\mtx_1,\mtx_2)$:
\begin{align}\label{eq5.54}
    &\psi_\mathrm{KG} \equiv \psi_\mathrm{KG}(\mtx_1, \mtx_2) = \frac{1}{\sqrt{2}} \left( \psi_1(\mtx_1) \psi_2(\mtx_2) + \psi_1(\mtx_2) \psi_2(\mtx_1) \right) 
\end{align}
The wavefunction Eq.\ (\ref{eq5.54}) satisfies two Klein-Gordon equations:
\begin{align}
    g^{\mu\nu} D_{1\mu} D_{1\nu} \psi (\mtx_1, \mtx_2) &= 0 
    \vt \\
    g^{\mu\nu} D_{2\mu} D_{2\nu} \psi(\mtx_1, \mtx_2) &= 0 
    \vt 
\end{align}
for each pair of spacetime variables $\mtx_1, \mtx_2$. Following the prescription of the optical approximation elucidated in the main text, these reduce to 
\begin{align}
    \p_{1\mu} \p^\mu_1 \psi(\mtx_1 , \mtx_2 ) &= 0 
    \vt 
    \\
    \p_{2\mu} \p^\mu_2 \psi(\mtx_1, \mtx_2 ) &= 0 
    \vt 
\end{align}
which admit plane wave solutions in the radial direction. One thus obtains two conserved currents, 
\begin{align}
    j^\mu_1 &= 2 \mathrm{Im} \: \psi^\star( \mtx_1, \mtx_2 ) \p^\mu_1 \psi( \mtx_1, \mtx_2 ) 
    \vt 
    \label{eq77}
    \\
    j^\mu_2 &= 2 \mathrm{Im} \: \psi^\star(\mtx_1, \mtx_2) \p^\mu_2 \psi(\mtx_1, \mtx_2 ) 
    \vt 
    \label{eq78}
\end{align}
which individually satisfy a continuity equation, 
\begin{align}\label{eq5.43}
    \p_{\mu 1} j^\mu_1 &= 0 
    \vt 
    \\
    \label{eq5.44}
    \p_{\mu 2} j^\mu_2 &= 0 
    \vt 
\end{align}
To obtain the velocity field of each particle, one uses the Bohmian-type formulae (as motivated by our single-particle derivation), 
\begin{align}
\label{eq31}
    v_1(\mtx_1, \mtx_2) &= f(r_1) \frac{j_1^1(\mtx_1, \mtx_2)}{j_0^0(\mtx_1, \mtx_2)} 
    \\
\label{eq32}
    v_2( \mtx_1, \mtx_2) &= f(r_2) \frac{j_2^1(\mtx_1, \mtx_2)}{j_2^0(\mtx_1, \mtx_2) }
\end{align}
where each velocity field is formed from the components of the Klein-Gordon conserved current vector, $\mathbf{j}_i \equiv (j^0_i, j^\mu_i) = ( \rho_i , j_i ) $ for particle $i = 1, 2$. Motivated by our weak value formalism, we evaluated trajectories assuming that both photons evolve on a single timeslice:
\begin{align}\label{eq21}
    v_{1,t} &\equiv v_{1,t} (t,r_1^\star,r_2^\star) = v_1 (\mtx_1, \mtx_2) \big|_{t_1 = t_2 = t}
    \vt 
    \\
    \label{eq22}
    v_{2,t} &\equiv v_{2,t} (t,r^\star_1,r^\star_2) = v_2 ( \mtx_1, \mtx_2) \big|_{t_1 = t_2 = t }
    \vt 
\end{align}
We re-emphasise that though this is the simplest choice for evaluating the trajectories, we are not constrained by it. Operationally, it is a natural way of viewing an experiment performed in a particular reference frame with detectors tracking the evolution of each photon trajectory by performing successive detections on an ensemble. We are additionally motivated by the possibility of a field-theoretic formulation of our measurement-based trajectories, in which the standard approach (i.e.\ in QFT) uses a single time coordinate \cite{mandel1995optical}.

\vspace{10pt}

With the preceding material in mind, we are now able to demonstrate the equivalence of the two approaches. First, we notice that the weak value velocity fields can be simplified as follows:

\begin{align}
    v_1 (t,r_1^\star,r_2^\star) &= f(r_1) \frac{2 \mathrm{Re} \: \langle \psi(t) | \bar{r} \rangle \langle \bar{r} | \hat{\mathbf{p}}_A | \psi(t) \rangle}{2 \mathrm{Re} \: \langle \psi(t) | \bar{r} \rangle \langle \bar{r} | \hat{\mathbf{H}}_A | \psi(t) \rangle}
    \\
    v_2 (t,r^\star_1,r^\star_2) &= f(r_2) \frac{2 \mathrm{Re} \: \langle \psi(t) | \bar{r} \rangle\langle \bar{r} | \hat{\mathbf{p}}_B | \psi(t) \rangle}{2 \mathrm{Re} \: \langle \psi(t) | \bar{r} \rangle \langle \bar{r} | \hat{\mathbf{H}}_B | \psi(t) \rangle }
\end{align}
In this form, it can be straightforwardly shown that the components of the Klein-Gordon conserved current, Eq.\ (\ref{eq77}) and (\ref{eq78}), are equivalent to these:
\begin{align}
    j_1^1(\mtx_1,\mtx_2) \big|_{t_1 = t_2 = t} &= 2 \mathrm{Re} \: \langle \psi(t) | \bar{r} \rangle\langle \bar{r} | \hat{\mathbf{p}}_A | \psi(t) \rangle  
    \vt 
    \\
    j_2^1( \mtx_1, \mtx_2) \big|_{t_1 = t_2 = t} &= 2\mathrm{Re} \: \langle \psi(t) | \bar{r} \rangle\langle \bar{r} | \hat{\mathbf{p}}_B | \psi(t) \rangle 
    \vt 
    \\
    \label{eq46}
    j_1^0(\mtx_1,\mtx_2) \big|_{t_1 = t_2 = t} &= 2 \mathrm{Re} \: \langle \psi(t) | \bar{r} \rangle \langle \bar{r} | \hat{\mathbf{H}}_A | \psi(t) \rangle 
    \vt 
    \\
    \label{eq47}
    j_2^0( \mtx_1, \mtx_2) \big|_{t_1 = t_2 = t} &= 2 \mathrm{Re} \: \langle \psi(t) | \bar{r} \rangle \langle \bar{r} | \hat{\mathbf{H}}_B | \psi(t) \rangle 
    \vt
\end{align}
implying
\begin{align}
    v_{i\mathrm{W}}(t,x_1,x_2) &= v_{i\mathrm{KG}}(\mtx_1,\mtx_2) \big|_{t_1 = t_2 = t} 
\end{align}
where $i = 1,2$ and we have specially denoted the velocity fields as derived from the measurement-based and Klein-Gordon approaches by $v_{i\mathrm{W}}$ and $v_{i\mathrm{KG}}$ respectively.

\subsection{Components of the Klein-Gordon Current}

\noindent In this section, we state explicitly the components of the Klein-Gordon current used to integrate the two-photon velocity fields, resulting in the plots in Fig.\ \ref{fig:trajectories}. We assume that the ingoing and outgoing photons have equal central frequency $k_0$ and bandwidth $\sigma$, giving, 
\begin{align}
    \psi_1(t,r^\star) &= \left( \frac{2\sigma^2}{\pi} \right)^{1/4} \exp \Big[ - (t- r^\star) \left( i k_{0} + (t - r^\star ) \sigma^2 \right) \Big] 
    \\
    \psi_2(t,r^\star) &= \left( \frac{2\sigma^2}{\pi} \right)^{1/4} \exp \Big[ - (t + x) \left( i k_{0} + (t + r^\star) \sigma^2 \right) \Big] 
\end{align}
The position- and time-symmetrised wavefunction is thus,
\begin{align}
    \psi_\mathrm{KG}(\mtx_1, \mtx_2) &= \frac{\sigma}{\sqrt{\pi}} \Big( \exp \Big[ - V_2 ( i k_0 + V_2 \sigma^2 ) - U_1 ( i k_0 + U_1 \sigma^2 ) \Big] + \exp \Big[ - V_1 ( i k_0 + V_1 \sigma^2 ) - U_2 ( i k_0 + U_2 \sigma^2 ) \Big] \Big) 
\end{align}
where we have defined the coordinates $U_i = t_i - r_i^\star, V_i = t_i + r_i^\star$. The Klein-Gordon current densities are given by,
\begin{align}
    j_1^0 (\mtx_1,\mtx_2) \non \vphantom{\sqrt{\frac{2}{\pi}}}
    &= \frac{2\sigma^2k_0}{\pi} \big( \exp \left[ - 2 \sigma^2 \left( V_1^2 + U_2^2 \right) \right] + \exp \left[ - 2\sigma^2 ( V_2^2 + U_1^2) \right]
    \non \vphantom{\sqrt{\frac{2}{\pi}}}
    \\
    &  \qquad   + 2 \exp \left[ - \left( V_1^2 + V_2^2 \right) \sigma^2 - \left( U_1^2 + U_2^2 \right) \sigma^2 \right] \cos \big[ k_0 (U_1 - U_2 - V_1 + V_2) \big] \vphantom{\sqrt{\frac{2}{\pi}}}
    \non 
    \\
    &  \qquad   + \frac{2 \sigma^2}{k_0} \exp \left[ - \left( V_1^2 + V_2^2 \right) \sigma^2 - \left( U_1^2 + U_2^2 \right) \sigma^2 \right] (V_1 - U_1 ) \sin \big[ k_{0} (U_1 - U_2 - V_1 + V_2) \big] \vphantom{\sqrt{\frac{2}{\pi}}}
    \label{eq101}
    \\
    j_2^0 (\mtx_1,\mtx_2) \non \vphantom{\sqrt{\frac{2}{\pi}}}
    &= \frac{2\sigma^2k_0}{\pi} \big( \exp \left[ - 2\sigma^2 \left( V_2^2 + U_1^2 \right) \right]  + \exp \left[ - 2\sigma^2 ( V_1^2  + U_2^2 \right] \vphantom{\sqrt{\frac{2}{\pi}}}
    \non 
    \\
    &  \qquad  + 2 \exp \left[ - \left( V_1^2 + V_2^2 \right) \sigma^2- \left( U_1^2 + U_2^2 \right) \sigma^2 \right]  \cos \big[k_0(U_1 - U_2  - V_1 + V_2) \big] \vphantom{\sqrt{\frac{2}{\pi}}}
    \non 
    \\
    &  \qquad  - \frac{2 \sigma^2}{k_0} \exp \left[ - \left( V_1^2 + V_2^2 \right) \sigma^2 - \left( U_1^2 + U_2^2 \right) \sigma^2 \right] ( V_2 - U_2 ) \sin \big[ k_0 (U_1 - U_2- V_1 + V_2) \big] \vphantom{\sqrt{\frac{2}{\pi}}} 
\end{align}
The associated Klein-Gordon currents are
\begin{align}
    j_1^1(\mtx_1, \mtx_2) \non \vphantom{\sqrt{\frac{2}{\pi}}}
    &= \frac{2\sigma^2k_0}{\pi} \big( \exp \left[ - 2 \sigma^2 \left( V_2^2 + U_1^2 \right) \right] - \exp \left[ -2 \sigma^2 \left( V_1^2 + U_2^2 \right) \right] \vphantom{\sqrt{\frac{2}{\pi}}}
    \non 
    \\
    & \qquad + \frac{2 \sigma^2}{k_0} \exp \left[ - \left( V_1^2 + V_2^2 \right) \sigma^2 - \left( U_1^2 + U_2^2 \right) \sigma^2 \right] \left( V_1+ U_1 \right) \sin \big[ k_{0} (V_1 - V_2 - U_1 + U_2 ) \big] \big) \vphantom{\sqrt{\frac{2}{\pi}}} 
    \label{eq102}
    \\
    j_2^1(\mtx_1, \mtx_2 ) \non \vphantom{\sqrt{\frac{2}{\pi}}}
    &= \frac{2\sigma^2k_0}{\pi} \big( \exp \left[ - 2\sigma^2 \left( V_1^2 + U_2^2 \right) \right] - \exp \left[ - 2 \sigma^2 \left( V_2^2 + U_1^2 \right) \right]
    \non 
    \vphantom{\sqrt{\frac{2}{\pi}}}
    \\
    & \qquad - \frac{2 \sigma^2}{k_0} \exp \left[ - \left( V_1^2 + V_2^2 \right) \sigma^2 - \left( U_1^2 + U_2^2 \right) \sigma^2 \right] \left( V_2  + U_2 \right) \sin \big[ k_{0} ( V_1 - V_2 -  U_1 + U_2 ) \big] \big)
    \vphantom{\sqrt{\frac{2}{\pi}}}
\end{align}
Using these expressions to construct the Klein-Gordon velocity fields, one can solve the resulting implicit integral equations $\D r_i^\star/\D t = v_i(\mathbf{R}_1, \mathbf{R}_2)$ numerically, yielding the trajectory plots in the main text.

\end{widetext}

\end{document}